\documentclass[seceq]{ptptex}
\notypesetlogo

\usepackage{graphicx}
\usepackage{wrapfig}
\usepackage{wrapft}

\title{%
Non-equilibrium critical relaxation of the 3D Heisenberg magnets with long-range correlated disorder}

\author{%
Pavel V. \textsc{Prudnikov}\footnote{E-mail: prudnikp@univer.omsk.su},
Maria A. \textsc{Medvedeva},%
}

\inst{%
Dept. of Theoretical Physics, Omsk State University, Omsk 644077, Russia
}

\abst{%
Monte Carlo simulations of the short-time dynamic behavior are
reported for three-dimensional Heisenberg model with long-range
correlated disorder at criticality, in the case corresponding to
linear defects. The static and dynamic critical exponents are
determined for systems starting from an ordered initial state.
The obtained values of the exponents are
in a good agreement with results of the field-theoretic description
of the critical behavior of this model in the two-loop
approximation.}

\begin{document}

\maketitle


\section{Introduction}

Critical properties of disordered systems with short-range (SR) and
long-range (LR) correlated randomness have been studied extensively
\cite{Harris,Harris2,Khmelnitskii,Emery,Grinstein,Dorogovtsev,WeinribHalperin,Lawrie,Stinchcombe,Cardy,Folk}.
One important question to address is whether the introduction of
weak randomness changes the universality class of phase transition.
According to the well-known Harris criterion \cite{Harris} disorder
with SR correlations is relevant if $2 - d\nu_0 = \alpha_0 > 0$,
where $d$ is the spatial dimension, and $\nu_0$ and $\alpha_0$ are
the correlation-length and the specific-heat exponents of the pure
system. This criterion is modified in the presence of LR
correlations in the disorder. A special type of such a disorder has
been considered by Weinrib and Halperin (WH) \cite{WeinribHalperin}.
They showed that the disorder with power law correlation $g(x) \sim
x^{-a}$ for large separations $x$ is relevant if $2 - a\nu_0 > 0$
for $a < d$, whereas the usual SR Harris criterion recovers for $a
\geq d$. As a result, the existence of LR correlations in the
disorder gives significant effect and wider class of disordered
systems, not only the three-dimensional diluted Ising model with
point-like uncorrelated defects, can be characterized by a new
universality class of critical behavior.

The power law decay for the impurity-impurity pair correlation
function $g(x)$ allows a direct geometrical interpretation. So, for
integer $a$ it corresponds to the lines (at $a=d-1$) or the planes
(at $a=d-2$) of impurities of random orientation
\cite{WeinribHalperin}. Moreover, non-integer $a$ is sometimes
treated in terms of impurities fractal dimension \cite{LR_4He}.
Therefore, the models with LR-correlated quenched defects have both
theoretical interest due to the possibility of predicting new types
of critical behavior in disordered systems and experimental interest
due to the possibility of realizing LR-correlated defects in the
$\mathrm{{}^4\mathop{He}}$ in aerogels\cite{LR_4He}, polymers
\cite{LR_polymer}, and disordered solids containing fractal-like
defects \cite{Korzhenevskii_PRB1998} or dislocations near the sample
surface \cite{LR_surface}.

The critical exponents were calculated in
Ref.~\cite{WeinribHalperin} in the one-loop approximation using a
double expansion in $\varepsilon = 4 - d \ll 1$ and $\delta = 4 - a
\ll 1$. The correlation-length exponent was evaluated in this linear
approximation as $\nu=2/a$ and it was argued that this scaling
relation is exact and also holds in higher order approximation. In
paper \cite{Prudnikov_PRB2000} a renormalization analysis of WH
model was carried out directly for the 3D systems in the next
two-loop approximation with the values of $a$ in the range $2\leq a
\leq 3$. The static and dynamic critical exponents were calculated
with the use of the Pad\'{e}-Borel summation technique. The results
obtained in \cite{Prudnikov_PRB2000} essentially differ from the
results evaluated by a double $\varepsilon, \delta$ - expansion in
Ref.~\cite{WeinribHalperin}. The comparison of calculated the
exponent $\nu$ values and ratio $2/a$ showed the violation of the
relation $\nu = 2/a$, supposed in \cite{WeinribHalperin} as exact.

Ballesteros and Parisi \cite{BallesterosParisi} have studied by
Monte Carlo means the critical behavior in equilibrium of the 3D
site-diluted Ising model with LR spatially correlated disorder, in
the $a=2$ case corresponding to linear defects. They have computed
the critical exponents of these systems with the use of the
finite-size scaling techniques and found that a $\nu$ value is
compatible with the analytical predictions $\nu=2/a$.

In our paper \cite{Prudnikov_PTP2007} the integrated Monte Carlo
simulations of the short-time dynamic behavior have been carried out
for 3D Ising and XY models with LR-correlated disorder at
criticality, in the case corresponding to linear defects. Both
static and dynamic critical exponents were determined for systems
starting separately from ordered and disordered initial states. The
obtained values of the exponents are in good agreement with results
of the field-theoretic description of the critical behavior of these
models in the two-loop approximation \cite{Prudnikov_PRB2000}.

Also, in paper \cite{Ivaneyko_Physica2008} authors have performed
extensive Monte Carlo simulations of critical statics of 3D Ising
model with LR-correlated disorder with linear defects. The
Swendsen-Wang algorithm was used alongside with a histogram
reweighting technique and the finite-size scaling analysis to
evaluate the values of critical exponents. It was shown that
obtained estimates for exponents differ from both previous numerical
simulations \cite{BallesterosParisi,Prudnikov_PTP2007} and results
of renormalization-group (RG) calculations
\cite{WeinribHalperin,Korzhenevskii_PRB1998,Prudnikov_PRB2000}.

The present paper is devoted to numerically investigations of
critical dynamics in short-time regime of 3D Heisenberg magnets with
LR-correlated defects. Insertion disorder with LR correlations or
extended defects must modify a conventional critical dynamics of
pure Heisenberg ferro- or antiferromagnets which is described by the
model \textrm{J} and the model \textrm{G} in the classification of
Hohenberg and Halperin \cite{HalperinHoenberg} and can lead to
relaxational critical dynamics described by the model \textrm{A}
with number components $n=3$ for order parameter. The results of
Monte Carlo study of the nonequilibrium behavior in the planar
magnetics described by 2D XY-model with quenched structural defects
can be evidence of it \cite{2dXYdiluted}. A significant changes in
the time dependence of the autocorrelation function have been
observed in the low-temperature phase due to localization of the
spin excitations on structural defects.

The method of short-time critical dynamics \cite{STDreview} gives
the possibility to determine both static and dynamic critical
exponents modified by LR correlations of impurities
\cite{Prudnikov_PTP2007}. In the following section, we introduce the
3D Heisenberg model with isotropic distributed linear defects and
scaling relations for the short-time critical dynamics. In
Sec.~\ref{sec:CM}, we derive the critical relaxation in short-time
regime for disordered Heisenberg systems starting from an ordered
initial state and cite values of static and dynamic critical
exponents obtained with the use of the leading corrections to
scaling. The final section contains analysis of the main results,
their comparison with results of other investigations, and our
conclusions.

\section{Description of the model and methods}

We have considered the following 3D site-diluted ferromagnetic
Heisenberg model Hamiltonian defined in a cubic lattice of linear
size $L$ with periodic boundary conditions:
\begin{eqnarray}
H =-J\sum_{\langle i,j\rangle}p_i p_j \vec{S}_i \vec{S}_j,
\end{eqnarray}
where $\vec{S}_i=(S_i^x,S_i^y,S_i^z)$, the sum is extended to the
nearest neighbors, $J>0$ is the short-range exchange interaction
between spins $\vec{S}_i$, and the $p_i$ are quenched random
variables ($p_i=1$, when the site $i$ is occupied by spin, and
$p_i=0$, when the site is empty), with LR spatial correlation. An
actual $p_i$ set will be called a sample from now on. We have
studied the next way to introduce the correlation between the $p_i$
variables for WH model with $a=2$, corresponding to linear defects.
We start with a filled cubic lattice and remove lines of spins until
we get the fixed spin concentration $p$ in the sample. We remove
lines along the coordinate axes only to preserve the lattice
symmetries and equalize the probability of removal for all the
lattice points. This model was referred in \cite{BallesterosParisi}
as the model with non-Gaussian distribution noise and characterized
by the isotropic impurity-impurity pair correlation function decays
for large $r$ as $g(r)\sim 1/r^2$.

In this paper we have investigated disordered Heisenberg magnets
with the spin concentrations $p = 0.8$. We have considered the cubic
lattices with linear size $L = 128$. The Metropolis algorithm has
been used in simulations. We consider only the dynamic evolution of
systems described by the model A in the classification of Hohenberg
and Halperin \cite{HalperinHoenberg}. The Metropolis Monte Carlo
scheme of simulation with the dynamics of a single-spin flips
reflects the dynamics of model \textrm{A} and enables us to compare
obtained critical exponents $z$ to the results of RG description of
the critical dynamics of this model\cite{Prudnikov_PRB2000}.

According to the argument of Janssen \textit{et al.} \cite{Janssen}
obtained with the RG method and $\varepsilon$-expansion, one may
expect a generalized scaling relation for the $k$th moment the
magnetization
\begin{equation} \label{mk}
\begin{split}
m^{(k)} &\left( t, \tau, L, m_0 \right)=b^{- k\beta / \nu } \\
        &\times m^{(k)}\left( b^{-z} t, \ b^{1/\nu} \tau, \ b^{-1}L, \ b^{x_0}m_0 \right)
\end{split}
\end{equation}
is realized after a time scale $t_{\rm mic}$ which is large enough
in a microscopic sense but still very small in a macroscopic sense.
In Eq.~(\ref{mk}), $b$ is a spatial rescaling factor, $\beta$ and
$\nu$ are the well-known static critical exponents, and $z$ is the
dynamic exponent, while the new independent exponent $x_0$ is the
scaling dimension of the initial magnetization $m_0$ and
$\tau=(T-T_c)/T_c$ is the reduced temperature.

The short-time dynamic method in part of critical evolution
description of system starting from the ordered initial state is
essentially the same as the non-equilibrium relaxation method
proposed by Ito in Ref.~\cite{Ito_Review,Ito} for non-equilibrium critical behavior study.

Since the system is in the early stage of the evolution the
correlation length is still small and finite size problems are
nearly absent. Therefore, we generally consider $L$ large enough
($L=128$) and skip this argument.
We measured the time evolution of the magnetization determined as
follows:
\begin{equation} \label{m1}
m(t)= \Bigg[\!\Bigg<
      \frac{1}{N_s}
      \bigg(  \big( \textstyle\sum\nolimits_{i}^{N_s} p_i S_{i}^{x} \big)^2 +
               \big( \textstyle\sum\nolimits_{i}^{N_s} p_i S_{i}^{y} \big)^2 +
               \big( \textstyle\sum\nolimits_{i}^{N_s} p_i S_{i}^{z} \big)^2
      \bigg)^{\!1/2} \, \Bigg>\!\Bigg]
\end{equation}
where angle brackets denote the statistical averaging, the square
brackets are for averaging over the different impurity
configurations, and $N_s=pL^3$ is a number of spins in the lattice.

The question arises how a completely ordered initial state with
$m_0=1$ evolves, when heated up suddenly to the critical
temperature. In the scaling form (\ref{mk}), one can skip besides
$L$, also the argument $m_0=1$,
\begin{equation}
m^{(k)}(t,\tau)=b^{-k\beta/\nu}m^{(k)}\left(b^{-z}t,b^{1/\nu}\tau \right).
\end{equation}

The system is simulated numerically by starting with a completely
ordered state, whose evaluation is measured at or near the critical
temperature. The quantities measured are $m(t)$ and $m^{(2)}(t)$.
With $b=t^{1/z}$, one avoids the main $t$ dependence in $m^{(k)}(t)$
and for $k=1$ one has
\begin{eqnarray} \label{moder}
m(t,\tau)&=&t^{-\beta/\nu z}m(1,t^{1/\nu z}\tau) \\
         &=& t^{-\beta/\nu z}\left(1+at^{1/\nu z}\tau+O(\tau^2)\right). \nonumber
\end{eqnarray}
For $\tau=0$, the magnetization decays by a power law $m(t)\sim t^{-\beta/\nu z}$.
If $\tau \neq 0$, the power law behavior is modified by the scaling function
$m(1,t^{1/\nu z} \tau)$. From this fact, the critical temperature $T_c$ and
the critical exponent $\beta/\nu z$ can be determined.

The scaling form of magnetization in Eq.~(\ref{moder}) is presented as follows:
\begin{equation} \label{logm}
\ln  m(t,\tau) = (-\beta/\nu z) \ln t + \ln m(1,t^{1/\nu z}\tau)
\end{equation}
after differentiation with respect to $\tau$ gives the power law of
time dependence for the logarithmic derivative of the magnetization
in the following form:
\begin{equation} \label{logderm}
\left.\partial_{\tau} \ln  m(t,\tau) \right|_{\tau=0} \sim t^{1/\nu z},
\end{equation}
which allows to determine the ratio $1/\nu z$. On the basis of the
magnetization and its second moment, the cumulant
\begin{equation} \label{bcum}
U_2(t)= \frac{m^{(2)}}{(m)^2} - 1 \sim t^{d/z}
\end{equation}
is defined. From its slope, one can directly measure the dynamic
exponent $z$. Consequently, from an investigation of the system
relaxation from ordered initial state with $m_0=1$, the dynamic
exponent $z$ and the static exponents $\beta$ and $\nu$ can be
determined.

The critical dynamic exponent $z$ of the model can be obtained also
from the time evolution of the ratio \cite{F2z}
\begin{equation}\label{F2}
F_2(t)=\frac{\left. m^{(2)}(t)\right|_{m_0=0}}
{\left. \left[ m(t)\right]^2\right|_{m_0=1}}
\sim \frac{t^{\displaystyle (d-2\beta/\nu)/z}}{t^{\displaystyle -2\beta/\nu z}}
=t^{\displaystyle d/z}.
\end{equation}

\section{Measurements of the critical exponents for 3D Heisenberg
model with linear defects\label{sec:CM}}

We have performed simulations on three-dimensional cubic lattices
with linear size $L=128$, starting from an ordered initial state. We
would like to mention that measurements starting from a completely
ordered state with the spins oriented in the same direction ($m_0 =
1$) are more favorable, since they are much less affected by
fluctuations, because the quantities measured are rather big in
contrast to those from a random start with with zero or small
initial magnetization ($m_0 \ll 1$). Therefore, for careful
determination of the critical temperature and critical exponents for
3D Heisenberg model with linear defects, we investigate the
relaxation of this model from a completely ordered initial state.

Initial configurations for systems with the spin concentration
$p=0.8$ and randomly distributed quenched linear defects were
generated numerically. Starting from those initial configurations,
the system was updated with Metropolis algorithm. Simulation have
been performed up to $t=1000$ Monte Carlo steps per spin (MCS/s).

We measured the time evolution of the magnetization $m(t)$ and the
second moment $m^{(2)}(t)$, which also allow to calculate the
time-dependent cumulant $U_2(t)$ in Eq.~(\ref{bcum}).

\begin{figure}[t]
\centering
\includegraphics[width=0.45\textwidth]{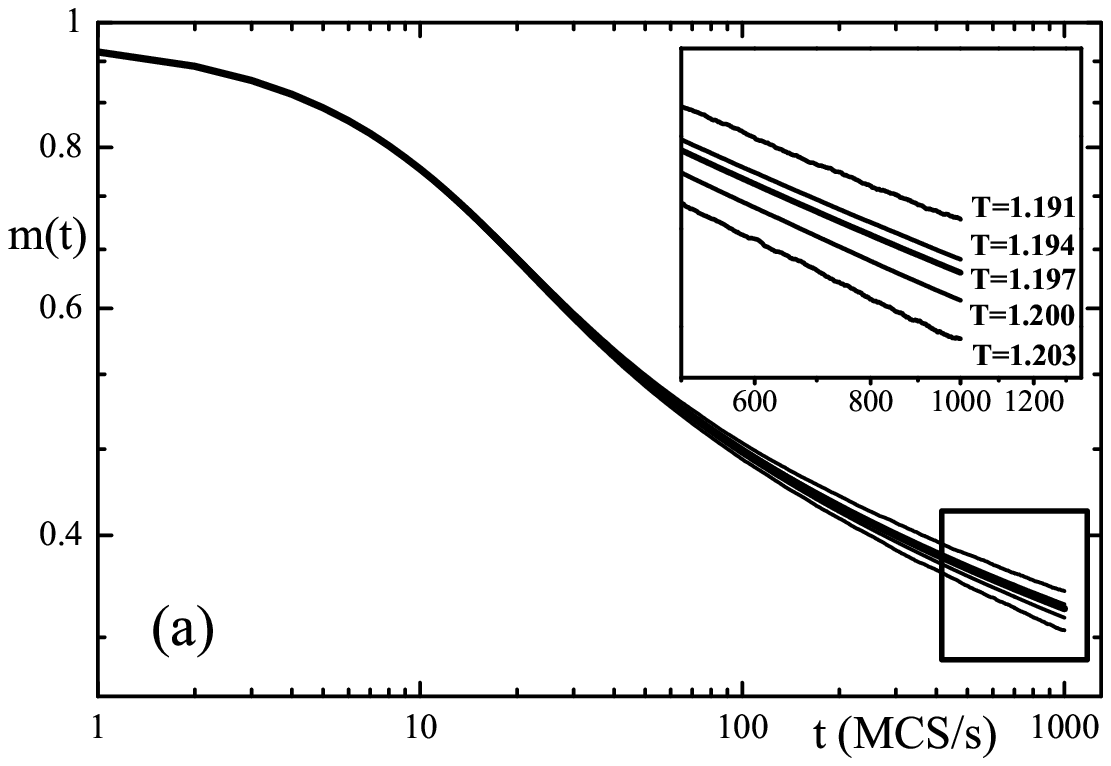}
\includegraphics[width=0.45\textwidth]{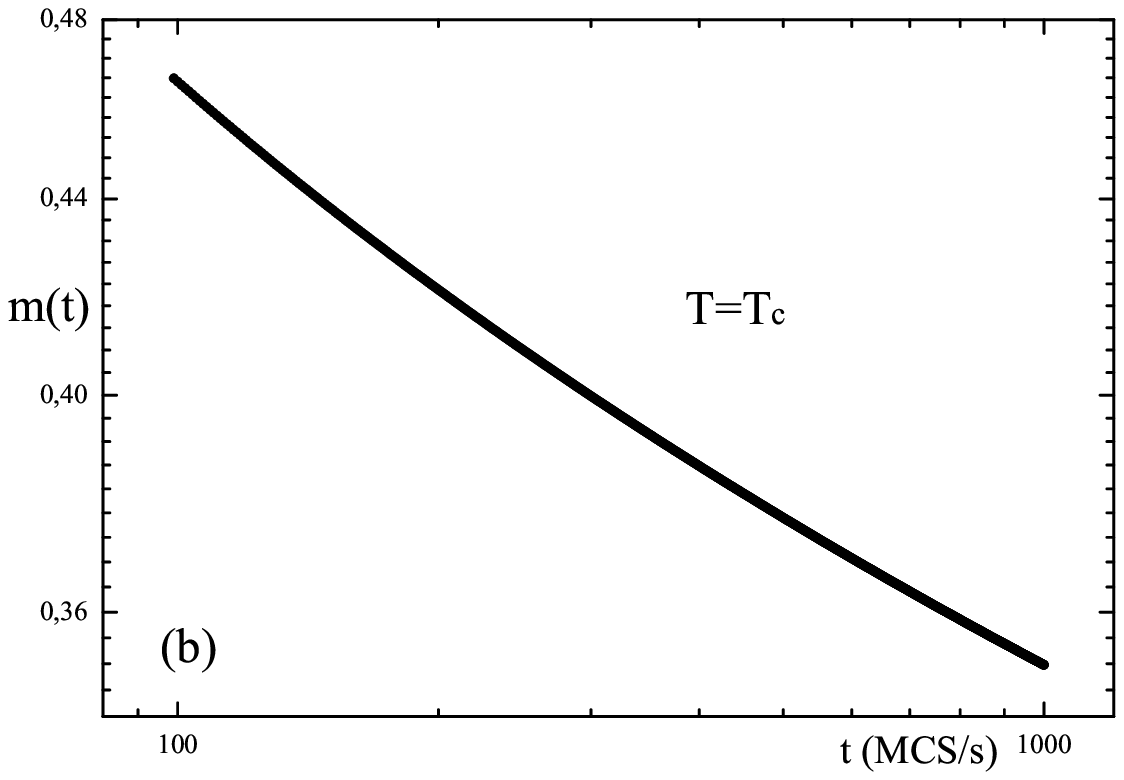}\\
\caption{\label{fig:m} Time evolution of the magnetization $m(t)$
 for different values of the temperature $T$ \textbf{(a)} and for critical temperature $T=T_c=1.197(2)$ \textbf{(b)}. }
\end{figure}

In Fig.~\ref{fig:m}a the magnetization $m(t)$ for samples with
linear size $L=128$ at $T=1.191$, $1.194$, $1.197$, $1.20$, and
$1.203$ is plotted in log-log scale.  The resulting curves in
Fig.~\ref{fig:m} have been obtained by averaging over $2800$ samples
with different linear defects configurations with $25$ runs for each
sample. We have determined the critical temperature $T_c = 1.197(2)$
from best fitting of these curves by power law. The magnetization
$m(t)$ at the critical temperature $T=T_c$ is plotted in
Fig.~\ref{fig:m}b.

In order to check-up the critical temperature value independently,
we have carried out in equilibrium the calculation of cumulant
$U_4$, defined as
\begin{equation}
U_4 = \frac{1}{2}\left( 3 -
\frac{[\left<m^4\right>]\phantom{^2}}{[\left<m^2\right>]^2}\right),
\end{equation}
and the correlation length \cite{CorrLength}
\begin{eqnarray}
\xi&=&\frac{1}{2\sin{(\pi/L)}}\sqrt{\frac{\chi}{F}-1\,}, \\
\chi&=&\frac{1}{N_{s}}[\langle M^2\rangle], \\
F&=&\frac{1}{N_{s}}[\langle\Phi\rangle],  \\
\Phi&=&\frac{1}{3}\sum_{k=\{x,y,z\}}\sum_{n=1}^{3}\left(\left|\sum_j{p_j{S_j^k}
\exp\left(\frac{2\pi ix_{n,j}}{L}\right)}\right|^2\right),
\end{eqnarray}
where $(x_{1,j},x_{2,j},x_{3,j})$ are coordinates of $j$-th site of
lattice.

\begin{figure}[t]
\centering
\includegraphics[width=0.45\textwidth]{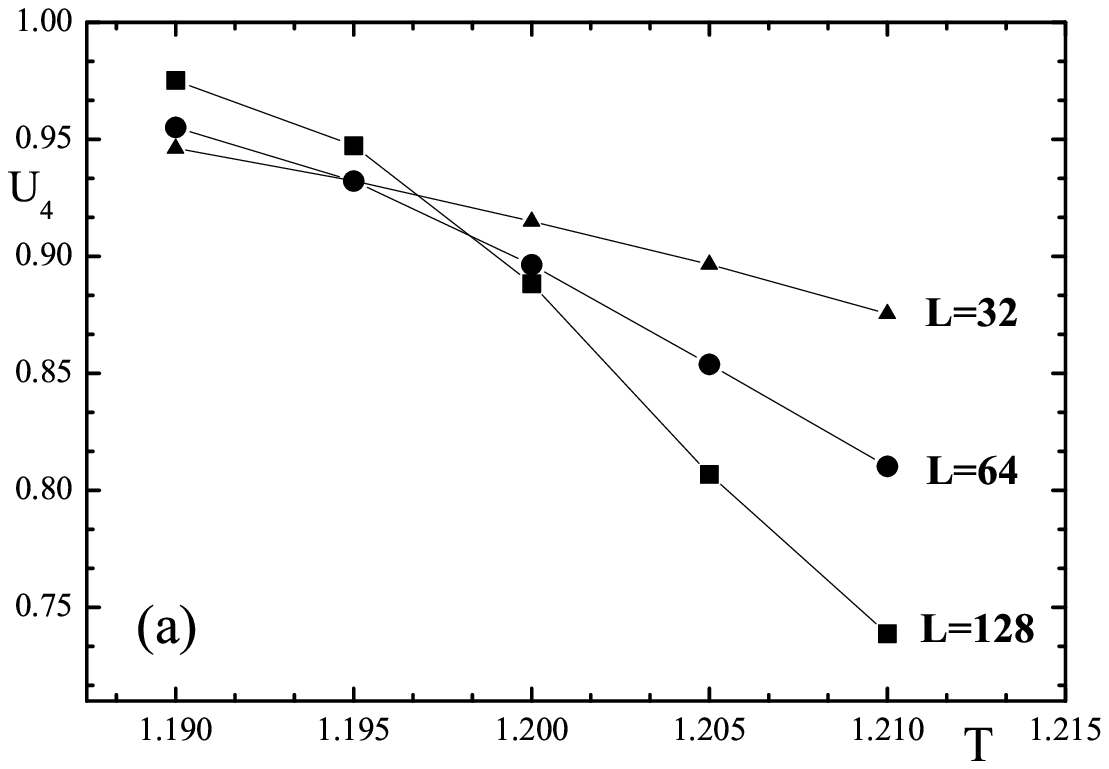}
\includegraphics[width=0.45\textwidth]{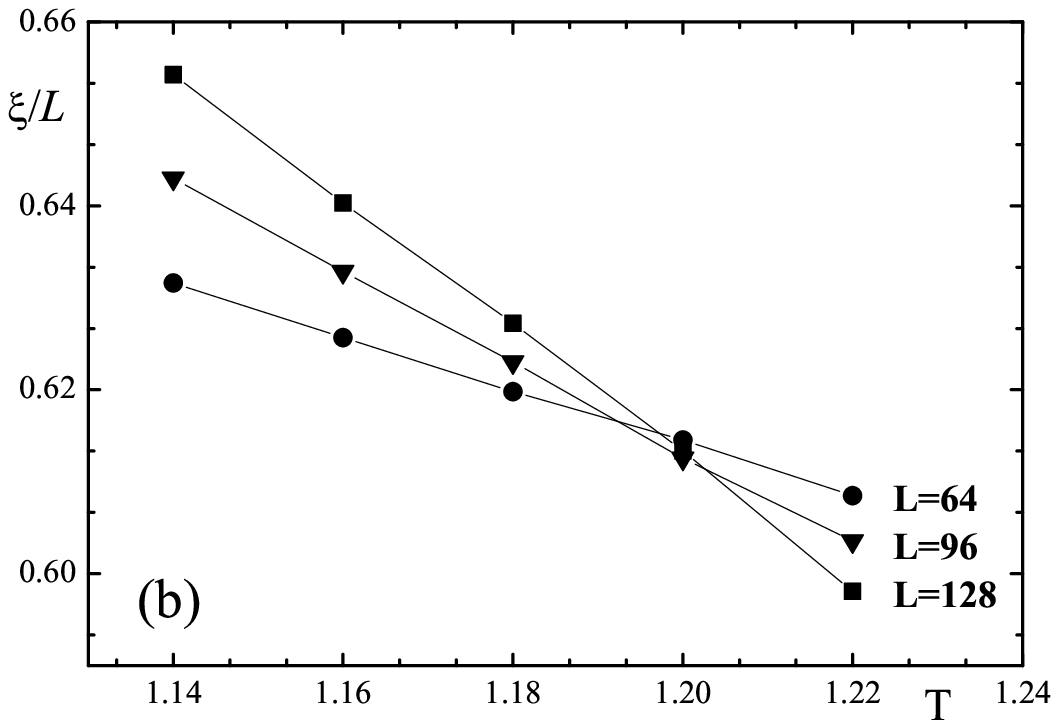}\\
\caption{\label{fig:Tc} Cumulant $U_4(T,L)$ (a) and ratio $\xi/L$
(b) as a function of $T$ for lattices with different sizes $L$. }
\end{figure}

The cumulant $U_4 (L,T)$ has a scaling form
\begin{equation}
U_4(L,T) = u\left(L^{1/\nu}(T-T_c)\right).
\end{equation}
The scaling dependence of the cumulant makes it possible to
determine the critical temperature $T_c$ from the coordinate of the
points of intersections of the curves specifying the temperature
dependence $U_4 (L,T)$ for different $L$. In Fig.~\ref{fig:Tc}a the
computed curves of $U_4(L,T)$ are presented for lattices with sizes
$L$ from $32$ to $128$. As a result it was determined that the
critical temperature is $T_c=1.197(2)$. In this case for simulations
we have used the Wolff single-cluster algorithm with elementary
MCS/s step as $5$ cluster flips. We discard $256$ MCS/s for
equilibration and then measure after every MCS with averaging over
$2048$ MCS/s. The results have been averaged over $1000$ different
samples with $25$ runs for each sample.

The crossing of $\xi/L$ was introduced as a convenient method for
calculating of $T_c$ in \cite{Tc_Ballesteros}. In Fig.~\ref{fig:Tc}b
the computed curves of temperature dependence of ratio $\xi/L$ are
presented for lattices with the same sizes, the coordinate of the
points of intersections of which also gives the critical temperature
$T_c=1.198(5)$. The value of $T_c=1.197(2)$ we selected as the best
for subsequent investigations of the Heisenberg model with linear
defects and with spin concentration $p=0.8$.

\begin{figure}[t]
\centering
\includegraphics[width=0.45\textwidth]{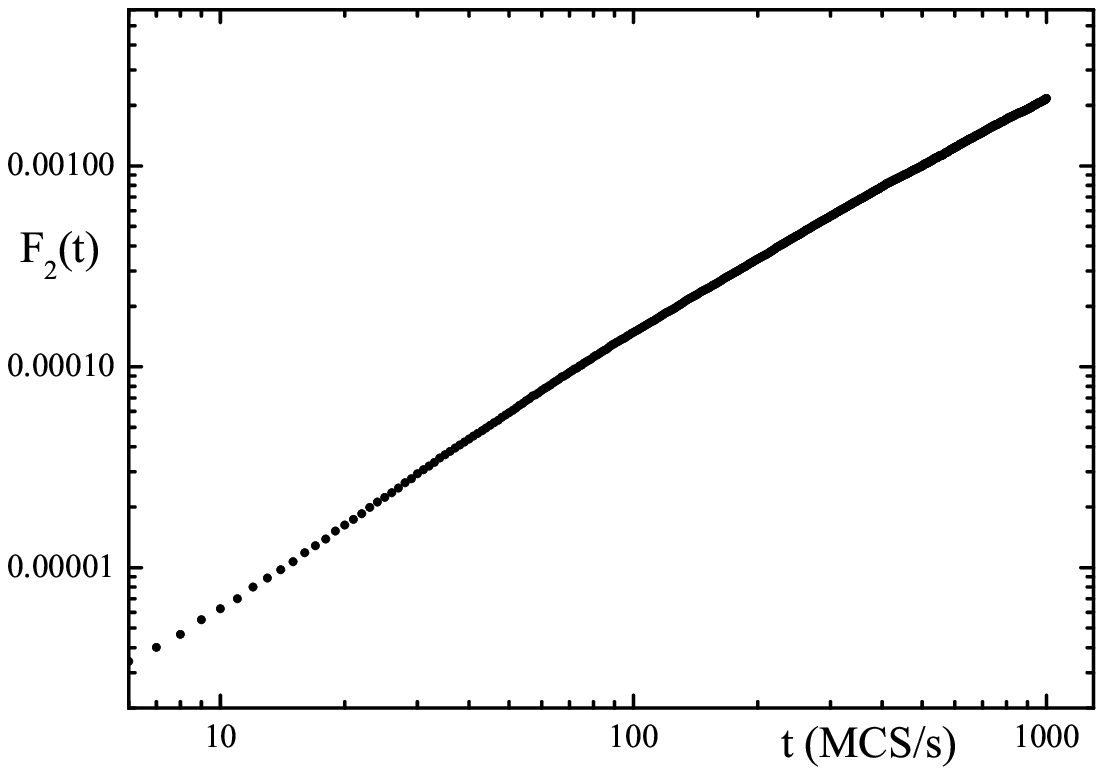}
\includegraphics[width=0.45\textwidth]{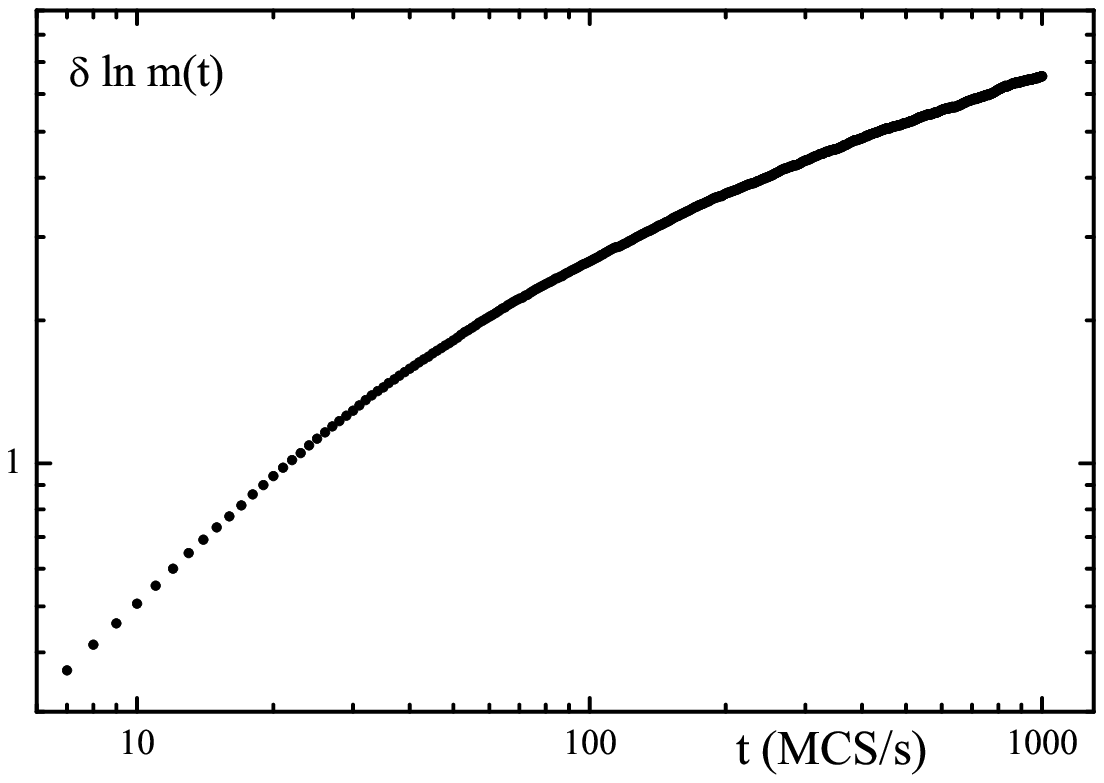}\\[-2mm]
\parbox[t]{0.45\textwidth}{
\caption{\label{fig:F2} Time evolution of the cumulant $F_2(t)$ is
plotted on a log-log scale at $T=T_c(p)$.}
}\quad
\parbox[t]{0.45\textwidth}{
\caption{\label{fig:LnM} Time evolution of the logarithmic derivative
of the magnetization $\left.\partial_{\tau} \ln  m(t,\tau)
\right|_{\tau=0}$ with respect to $\tau$ is plotted on a log-log scale.}
}
\end{figure}

In Fig.~\ref{fig:F2} the cumulant $F_2(t)$ and in
Fig.~\ref{fig:LnM}, the logarithmic derivative of the magnetization
$\left.\partial_{\tau} \ln  m(t,\tau) \right|_{\tau=0}$ with respect
to $\tau$ are plotted on a log-log scale at $T=T_c$. The
$\left.\partial_{\tau} \ln m(t,\tau) \right|_{\tau=0}$ have been
obtained from a quadratic interpolation between the three curves of
time evolution of the magnetization in Fig.~\ref{fig:m} for the
temperatures $T=T_c$, $T=T_c\mp 0.003$ and taken at the critical
temperature $T_c=1.197(2)$. The resulting curves in
Figs.~\ref{fig:m} and \ref{fig:F2} have been obtained at the
critical temperature by averaging over $3800$ samples with $25$ runs
for each sample.

In contrast to short-time dynamics of the pure systems \cite{Jaster}, we can
observe the crossover from dynamics of the pure system on early
times of the magnetization evolution from $t\simeq 15$ up to
$t\simeq 35$ MCS/s to dynamics of the disordered system with the
influence of long-range correlated defects for $t > 80$ MCS/s. The
same crossover phenomena were observed in the Monte Carlo simulated
critical short-time dynamic behavior of diluted 3D Ising systems with
point-like defects \cite{Prudnikov_PRE2010} and in behavior of 3D Ising
and XY systems with linear defects \cite{Prudnikov_PTP2007}.

\begin{figure}[t!]
\centering
\includegraphics[width=0.5\textwidth]{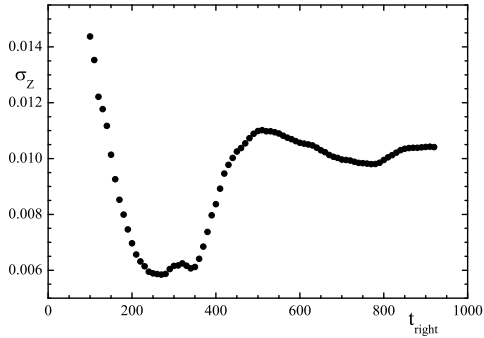}\\[-2mm]
\caption{\label{fig:Interval_Delta_z} Dependence of the mean square
error $\sigma_z$ as a function of the right border of the time
interval $t\in [80, t_{\text{\rm right}}]$ }
\end{figure}

The existence of different regimes in short-time dynamic evolution
of disordered system can be demonstrated clearly on critical time
behavior of the ratio $F_2(t)$ (Fig.~\ref{fig:F2}). We have analyzed
the time dependence of the $F_2(t)$ in the time interval $t \in
[15,35]$ MCS/s where the $F_2(t)$ is best fitted by power law with
the exponent $d/z = 1.464(22)$. This value of $d/z$ gives the
dynamic exponent $z=2.049(31)$, corresponding to the pure $O(n=3)$
Heisenberg model \cite{Prudnikov_JETP2008}. An analysis of the
$F_2(t)$ slope measured in the interval $t\in [80,300]$ MCS/s shows
that the exponent $d/z = 1.217(3)$ which gives $z = 2.465(6)$.
The dependence of the mean square error $\sigma_z$ as a function of
the right border of the time interval $t\in [80, t_{\text{\rm
right}}]$ is presented in Fig.~\ref{fig:Interval_Delta_z}. The time
interval $[80, t_{\text{\rm right}}]$ for $t_{\text{\rm right}}=300$
gives the minimum of errors for exponent $z$ in comparison with
value $z= 2.591(10)$, which evaluated in the time interval $t\in
[80, 1000]$.

The same analysis of time dependences of the magnetization $m(t)$
and the logarithmic derivative of the magnetization
$\left.\partial_{\tau} \ln m(t,\tau) \right|_{\tau=0}$ in the
initial time interval gives the exponent values $\beta/\nu z =
0.249(1)$ and $1/\nu z = 0.692(15)$ with the use of $z=2.049(31)$ we
obtain the values of exponents $\nu=0.705(26)$, $\beta=0.360(9)$ and
$\beta/\nu = 0.510(10)$. These values are in a good agreement with
exponents $\nu=0.7048(30)$, $\beta=0.3636(45)$ and
$\beta/\nu=0.5158(102)$ obtained in Ref.~\cite{FerrenbergLandau} on
the base of high resolution Monte Carlo study of pure 3D Heisenberg
model critical behavior.

An analysis of the $U_2(t)$ slope measured in the interval $t\in
[80,300]$ MCS/s shows that the exponent $d/z = 1.170(136)$ which
gives $z = 2.564(368)$. The value of the dynamic exponent $z$
obtained from the ratio $F_2$ in Eq.~(\ref{F2}) is more preferred
because evolution of $F_2(t)$ is less fluctuated then $U_2(t)$, and,
therefore, a greatly large statistics is necessary for obtaining the
same quality results from measurements of $U_2(t)$ as is in the case
with $F_2(t)$. So, the slope of magnetization $m(t)$ and its
derivative $\left.\partial \ln m(t)\right|_{\tau=0}$ measured in the
interval $t\in [80,300]$ MCS/s provides the ratio of exponents
$\beta/\nu z=0.150(1)$ and $1/\nu z=0.483(22)$ which give
$\beta=0.311(16)$ and $\nu=0.840(47)$.

These values of exponents can be compared with results of the
field-theoretic description of the critical behavior of Heisenberg
model with LR-correlated defects in the two-loop approximation
$z=2.264$, $\nu=0.798$, and $\beta=0.384$ calculated in
Ref.~\cite{Prudnikov_PRB2000} for case of Heisenberg system with
linear defects when the correlation parameter $a=2$. The some
numerical differences of these values exceeding the limits of
statistical errors of simulation and numerical approximations should
not be discouraged since the obtained effective values of exponents
cannot be considered as final.

\begin{figure}[t]
\centering
\includegraphics[width=0.3\textwidth]{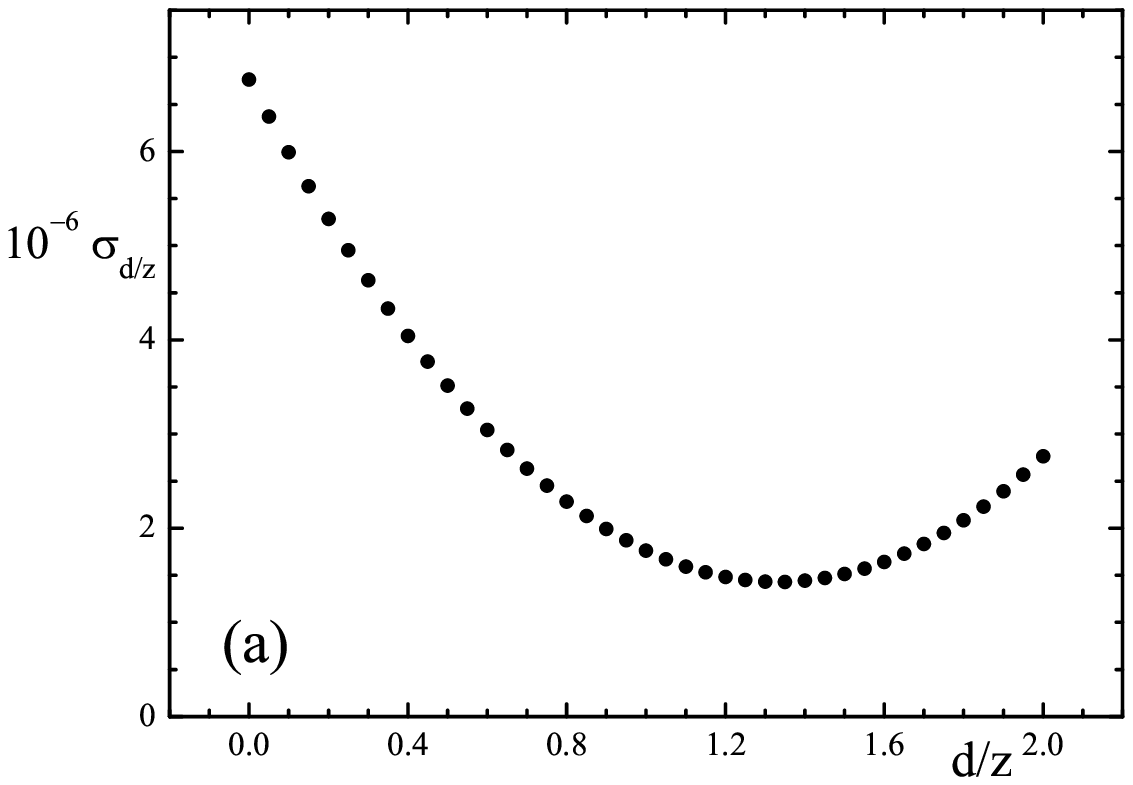}
\includegraphics[width=0.3\textwidth]{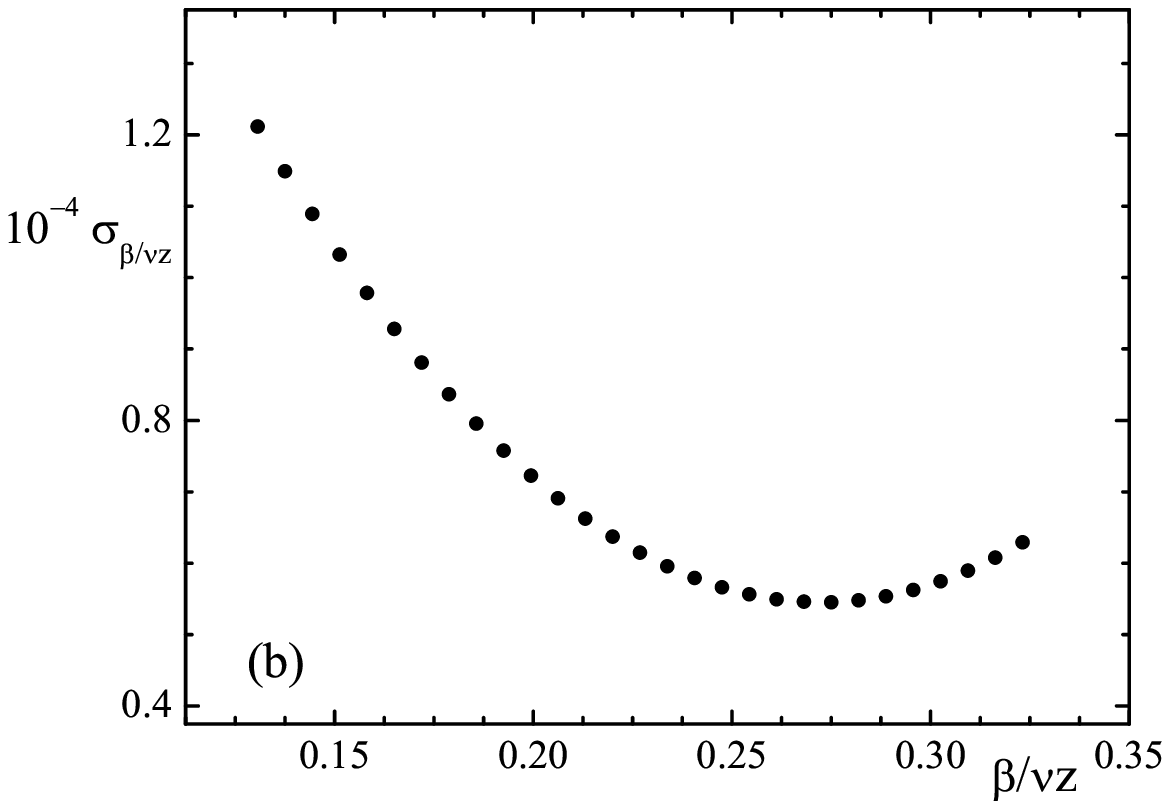}
\includegraphics[width=0.3\textwidth]{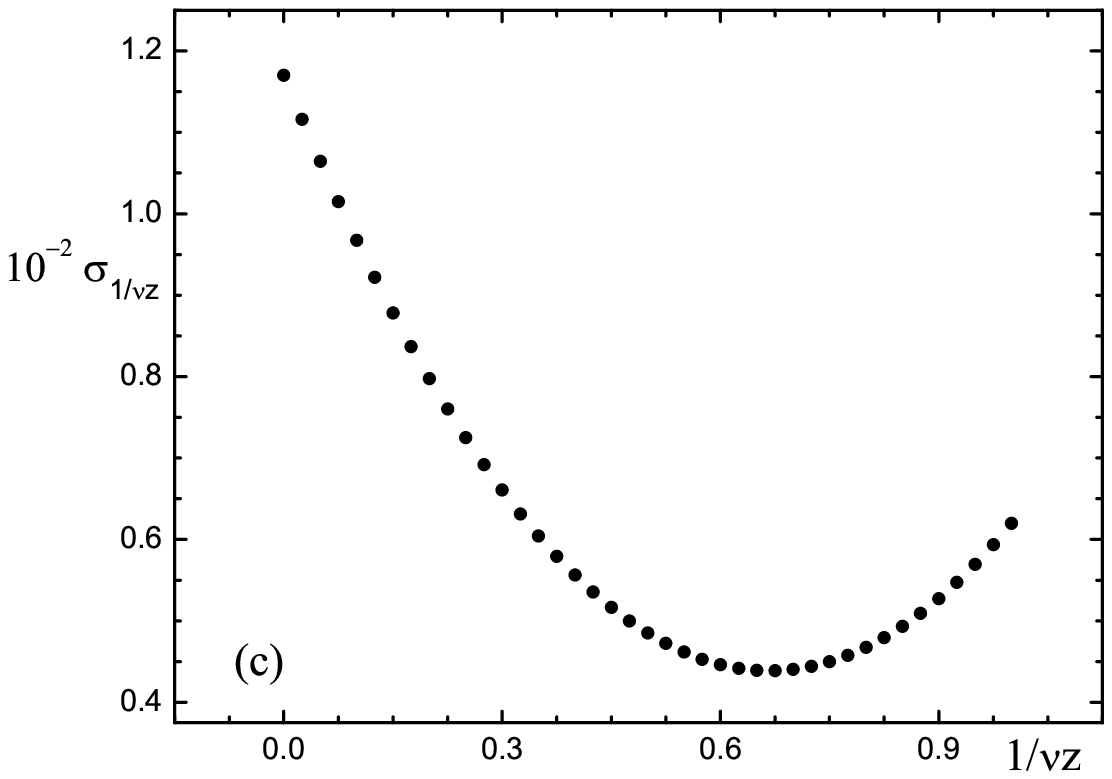}\\[-2mm]
\caption{\label{fig:sigma} Dependence of the mean-square errors
$\sigma$ of the fits for (a) the cumulant $F_2$, (b) magnetization
$m$, and (c) the logarithmic derivative of the magnetization
$\partial_{\tau} \ln  m(t,\tau)$ as a function of the exponents
$d/z$, $\beta/\nu z$ and $1/\nu z$ for
$\omega/z=(\omega/z)_\mathrm{min}$ and time interval $[80,300]$.}
\end{figure}

In the next stage in order to obtain accurate values of the critical
exponents, we have considered the influence of a leading corrections
to the scaling on asymptotic values of exponents. We have applied
the following expression for the observable $X(t)$:
\begin{eqnarray}\label{f_m}
X(t) = A_x t^{\delta}(1+B_x t^{-\omega/z}),
\end{eqnarray}
where $\omega$ is an exponent of the leading corrections to scaling,
$A_x$ and $B_x$ are fitting parameters, and an exponent
$\delta=-\beta/\nu z$ when $X\equiv m(t)$, $\delta=d/z$ when
$X\equiv F_2(t)$ or $X\equiv U_2(t)$, and $\delta=1/\nu z$ when
$X\equiv \left.\partial_{\tau} \ln  m(t,\tau)\right|_{\tau=0}$. The
expression in Eq.~(\ref{f_m}) reflects the scaling transformation in
the critical range of time-dependent corrections to scaling in the
form of $t^{-\omega/z}$ to the usual form of corrections to scaling
$\tau^{\omega \nu}$ in equilibrium state for time $t$ comparable to
the order parameter relaxation time $t_r \sim \xi^z \Omega(k\xi)$
\cite{HalperinHoenberg}.

\begin{figure}[t]
\centering
\includegraphics[width=0.3\textwidth]{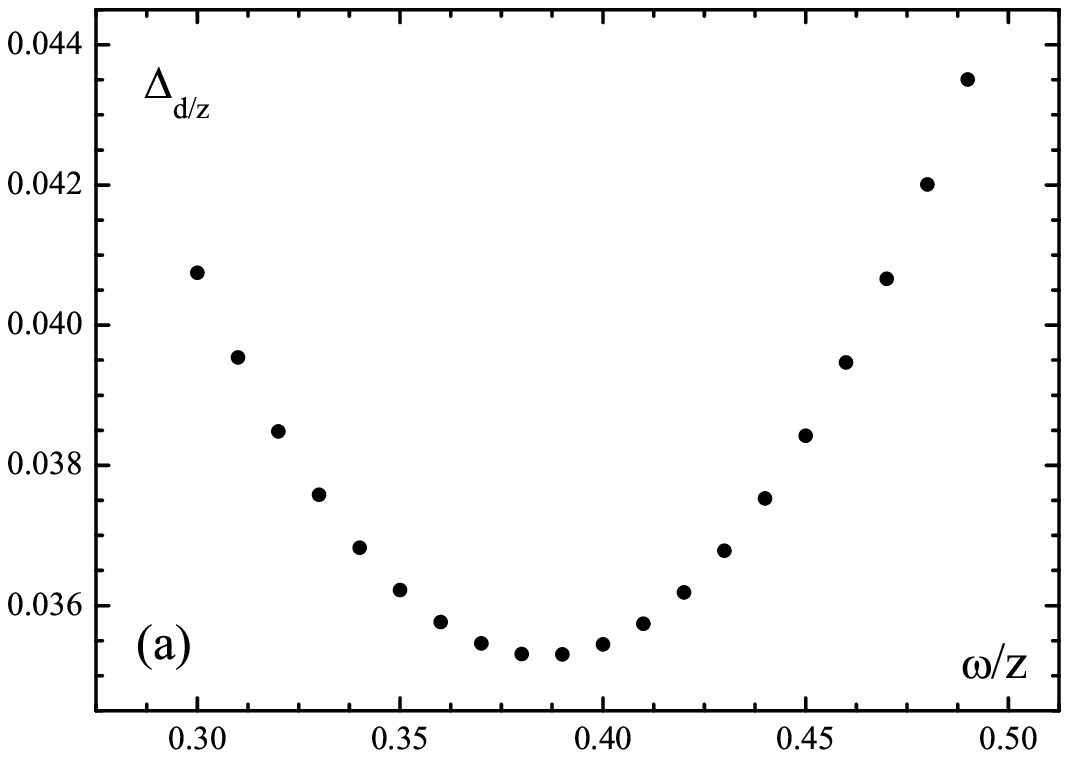}
\includegraphics[width=0.3\textwidth]{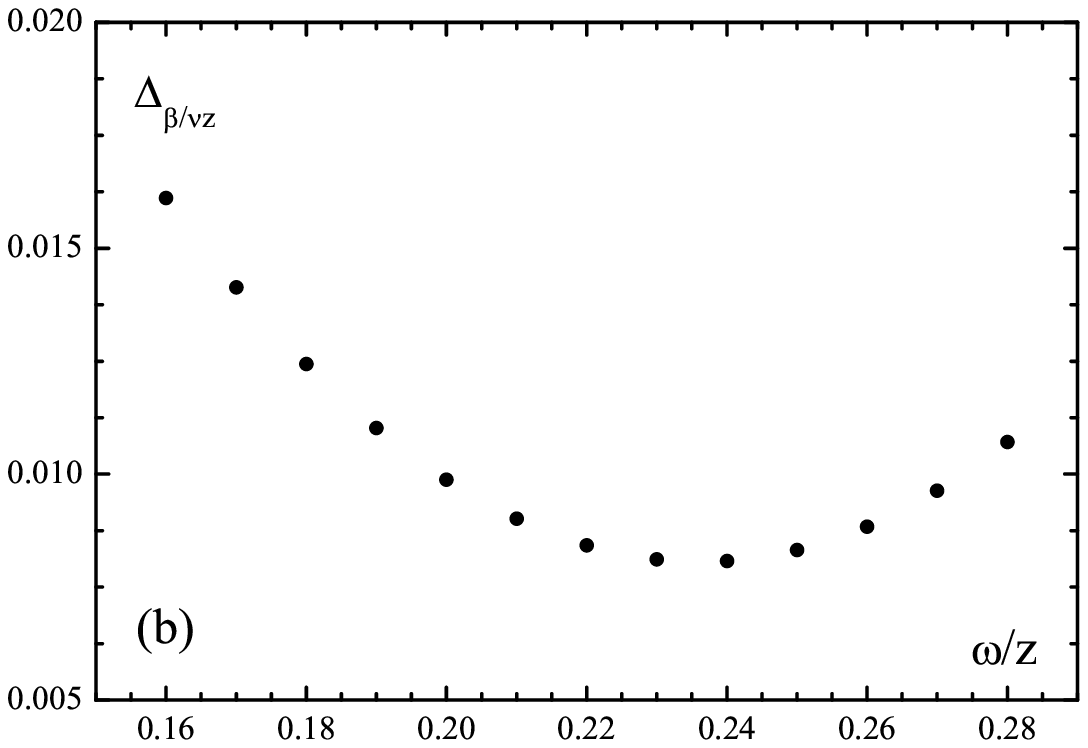}
\includegraphics[width=0.3\textwidth]{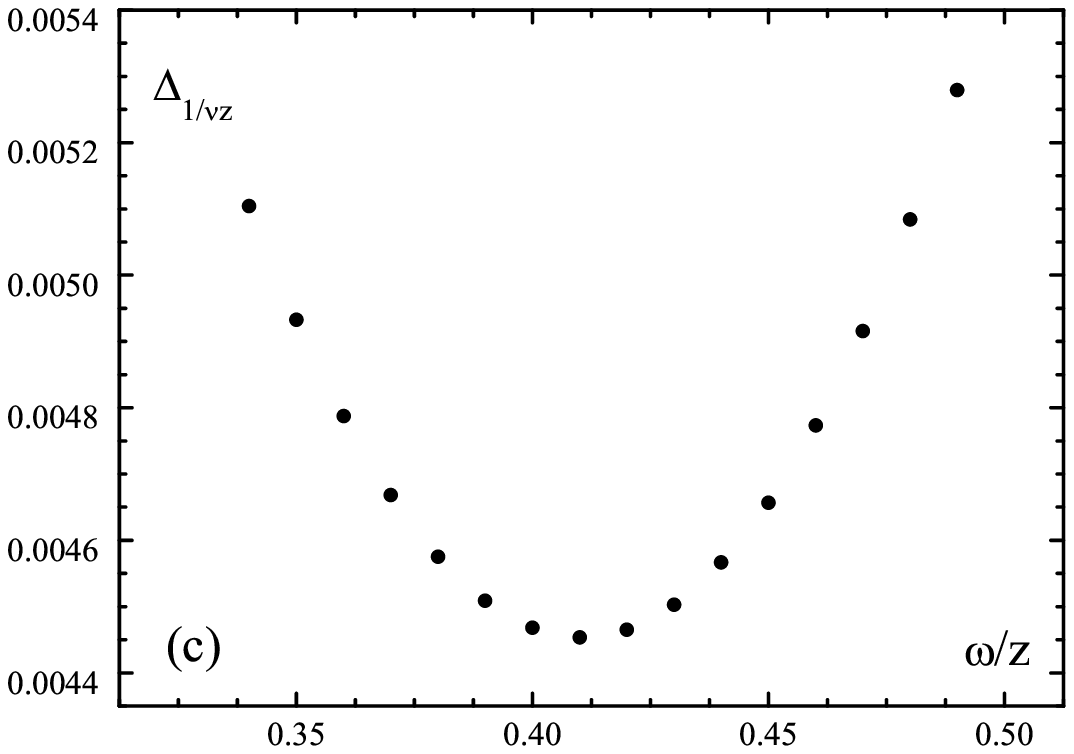}\\[-2mm]
\caption{\label{fig:Delta_dz} Dependence of global mean-square error
 (a) $\Delta_{d/z}$, (b) $\Delta_{\beta/\nu z}$ and (c) $\Delta_{1/\nu z}$  for all time intervals as a function of the
exponent $\omega/z$.}
\end{figure}

\begin{figure}[t!]
\centering
\includegraphics[width=0.7\textwidth]{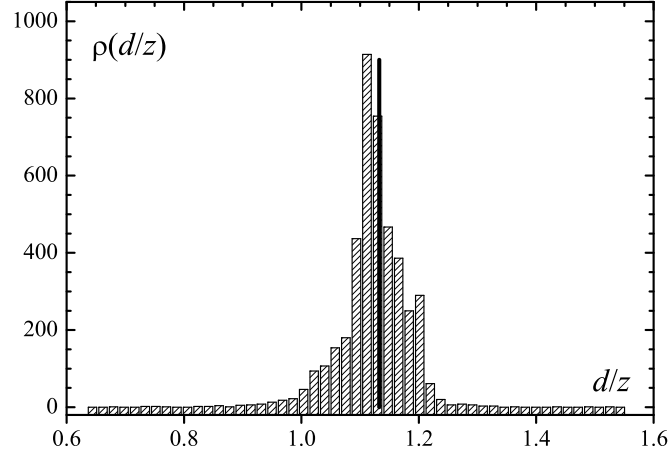}\\[-2mm]
\caption{\label{fig:rho} The distribution function $\rho(d/z)$ for
different time intervals. Solid line is corresponds to calculated averaged value $d/z=1.329$}
\end{figure}
\begin{figure}[t!]
\centering
\includegraphics[width=0.7\textwidth]{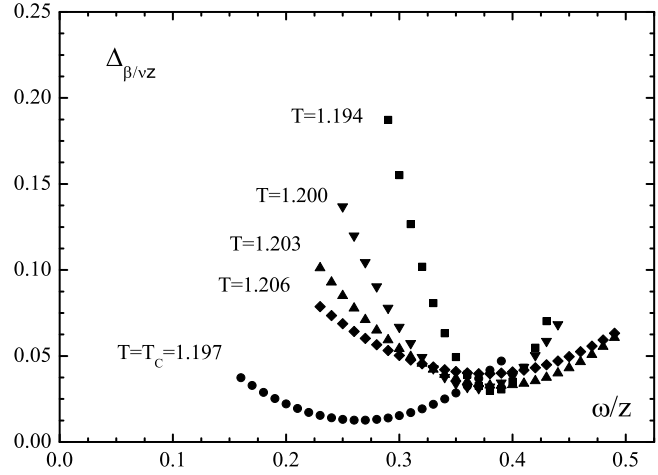}\\[-2mm]
\caption{\label{fig:Tc_Delta_dz} Dependence of global mean-square
error $\Delta_{\beta/\nu z}$ as a function of the exponent
$\omega/z$ for different values of $T$.}
\end{figure}

We have used the least-squares method for the best approximation of
the simulation data $X(t)$ by the expression in Eq.~(\ref{f_m}).
Minimum of the mean square errors $\sigma_\delta$ of this fitting
procedure determines the exponents $\delta$ and $\omega/z$. As
example, we plot in Fig.~\ref{fig:sigma} for time interval $[80,
300]$ the $\sigma_{d/z}$ (a) for the cumulant $F_2$ as a function of
the exponent $d/z$ for $\omega/z=0.40$, the $\sigma_{\beta/\nu z}$
(b) for the magnetization $m$ as a function of the exponent
$\beta/\nu z$ for $\omega/z=0.24$ and the $\sigma_{1/\nu z}$ (c) for
the logarithmic derivative of the magnetization as a function of the
exponent $1/\nu z$ for $\omega/z=0.41$. All obtained minimal values
of $\delta$ are averaged for all time intervals.
The  distribution function $\rho(\delta)$ for different time intervals is presented
in Fig.~\ref{fig:rho} for instance $\delta=d/z$. The global mean-square error
\begin{equation}
\Delta_{\delta}=\sqrt{\displaystyle
{\scriptstyle \sum_{i=1}^{N}{(\delta_i-\overline{\delta})^2\cdot \Delta t_i }}
\left/
{\scriptstyle \sum_{i=1}^{N}{ \Delta t_i } }
\right.
}
\end{equation}
is calculated on base of averaged values $\overline{\delta}$, where
$N$ is number of time intervals, which demonstrates minimum of
errors. The dependence of global mean-square error $\Delta_{d/z}$
(a) for the cumulant $F_2$, $\Delta_{\beta/\nu z}$ (b) for the
magnetization $m$ and $\Delta_{1/\nu z}$ (c) for the logarithmic
derivative of the magnetization as a functions of the exponent
$\omega/z$ are presented in Fig.~\ref{fig:Delta_dz}. The critical
exponents are calculated for $\omega/z$ which corresponds to minimum
of $\Delta_\delta$.

The value of exponent $d/z=1.336(35)$ was computed with
$\omega/z=0.40(1)$, value $\beta/\nu z=0.228(7)$ was computed with
$\omega/z=0.24(1)$ and value $1/\nu z=0.589(4)$ was computed with
$\omega/z=0.41(1)$. It can be determined the values of the critical
exponents $z=2.245(60)$, $\nu=0.757(26)$, $\beta=0.388(15)$. For the
averaged value $\omega/z=0.36(7)$ it was computed values of the
exponents $d/z=1.329(36)$, $\beta/\nu z=0.226(29)$, and $1/\nu
z=0.575(40)$. On the base of these values, we determine the final
values of the critical exponents $z=2.257(61)$, $\nu=0.770(74)$,
$\beta=0.393(77)$, and $\omega=0.786(45)$. The statistical errors
for exponents are estimated by dividing all data into five sets.

The dependences of global mean-square error $\Delta_{\beta/\nu z}$
as a function of the exponent $\omega/z$ for different values of $T$
are presented in Fig.~\ref{fig:Tc_Delta_dz} and they demonstrate
that temperature $T=1.197$, which was choosen as critical, gives the
minimal value of fitting errors.

\begin{table}[t]
\tabcolsep 2pt
\caption{\label{tab:all}Values of the obtained critical exponents
and comparison with results of renormalization group (RG) and Monte Carlo (MC)
calculations}
\begin{tabular}{ll|lllll} \hline\hline
                                                                                                                    && \multicolumn{1}{c}{$z$}  & \multicolumn{1}{c}{$\beta/\nu$} & \multicolumn{1}{c}{$\nu$} & \multicolumn{1}{c}{$\beta$} & \multicolumn{1}{c}{$\omega$}  \\ \hline
           $m_0=1$, LR system stage                                                                                 && $2.257(61)$ & $0.510(78)$   & $0.770(74)$ & $0.393(77)$ & $0.786(45)$  \\
           $m_0=1$, pure system stage                                                                               && $2.049(31)$ & $0.510(10)$   & $0.705(26)$ & $0.360(9)$ \\  \hline\hline
 \multicolumn{1}{c}{LR system ($a=2$)} \\ \hline
 Prudnikov, \emph{et al.}, 2000,       (Ref.~\cite{Prudnikov_PRB2000})      & RG $d=3$               & $2.264$     & $0.482$       & $0.798$      & $0.384$ \\
 Prudnikov, \emph{et al.}, 2010,       (Ref.~\cite{Prudnikov_OmSU2010})     & RG $d=3$               & $2.291(29)$ & $0.490(5)$    & $0.766(17)$  & $0.375(5)$ \\
 Blavats'ka, \emph{et al.}, 2001,      (Ref.~\cite{BlavatskaFerberHol})     & RG $d=3$               &             &               &              &             & $0.88$ \\ \hline\hline
 \multicolumn{1}{c}{Pure system} \\ \hline
 Prudnikov, \emph{et al.}, 2008,       (Ref.~\cite{Prudnikov_JETP2008})     &RG $\varepsilon$-exp.   & $2.020(7)$  &               &              &                    \\
 Guida,      \emph{et al.}, 1998,      (Ref.~\cite{GuidaZinnJustin})        &RG $d=3$                &             & $0.5178(13)$  & $0.7073(35)$ & $0.3662(27)$      \\
                                                                            &RG $\varepsilon$-exp.   &             & $0.5188(23)$  & $0.7045(55)$ & $0.3655(45)$      \\
 Chen, \emph{et al.}, 1993,            (Ref.~\cite{FerrenbergLandau})       &MC                      &             & $0.5158(102)$ & $0.7048(30)$ & $0.3636(45)$         \\ \hline\hline
\end{tabular}
\end{table}

\section{Analysis of results and conclusions}

The present results of Monte Carlo investigations allow us to
recognize that the short-time dynamics method is reliable for the
study of the critical behavior of the systems with LR-correlated
disorder and is the alternative to traditional Monte Carlo methods.
But in contrast to studies of the critical behavior of the pure
systems by the short-time dynamics method, in case of the systems
with quenched disorder corresponding to randomly distributed linear
defects after the microscopic time $t_{\rm mic}\simeq 10$ there
exist three stages of dynamic evolution. In the time interval of
$15-35$ MCS/s, the power-law dependences are observed in the
critical point for the magnetization $m(t)$, the logarithmic
derivative of the magnetization $\left.\partial_{\tau} \ln m(t,\tau)
\right|_{\tau=0}$, and the cumulant $F_2(t)$, which are similar to
that in the pure system. In the time interval $[80, 300]$, the
power-law dependences are observed in the critical point which are
determined by the influence of disorder. In the intermediate time
interval the crossover behavior is observed in the dynamic evolution
of the system. However, careful analysis of the slopes for dynamical
characteristics reveals that a correction to scaling should be
considered in order to obtain accurate results.

\begin{figure}[t!]
\centering
\hspace*{-3mm}\includegraphics[width=0.45\textwidth]{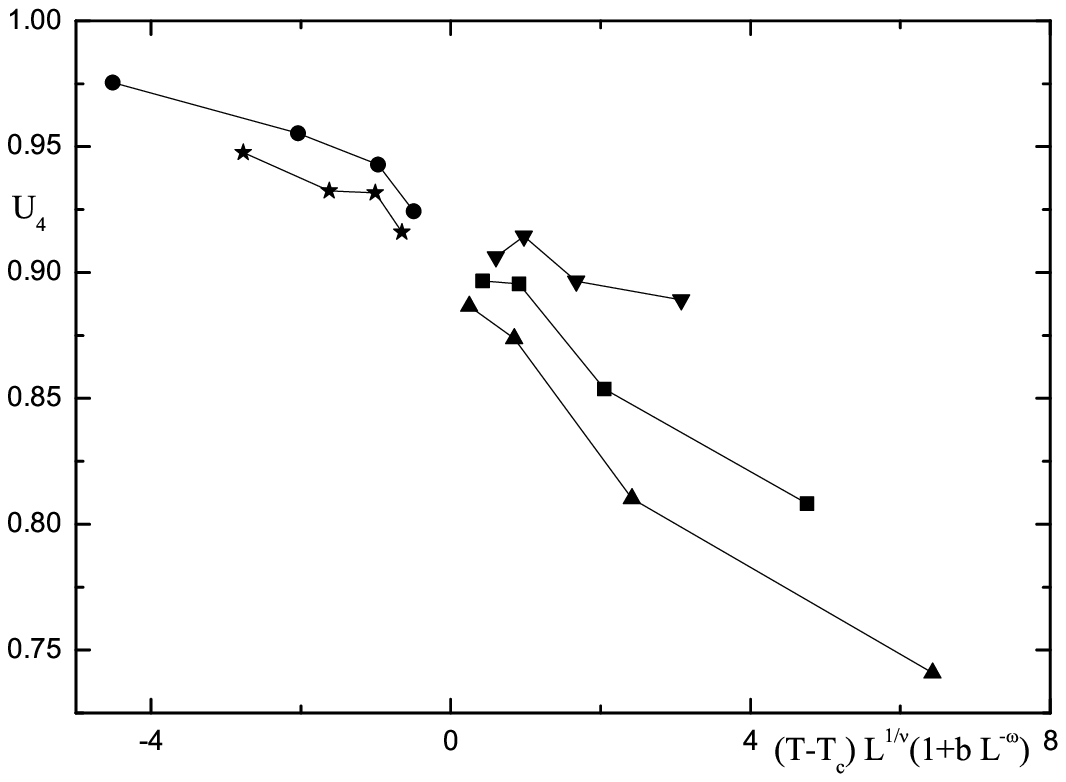}
\quad
\includegraphics[width=0.45\textwidth]{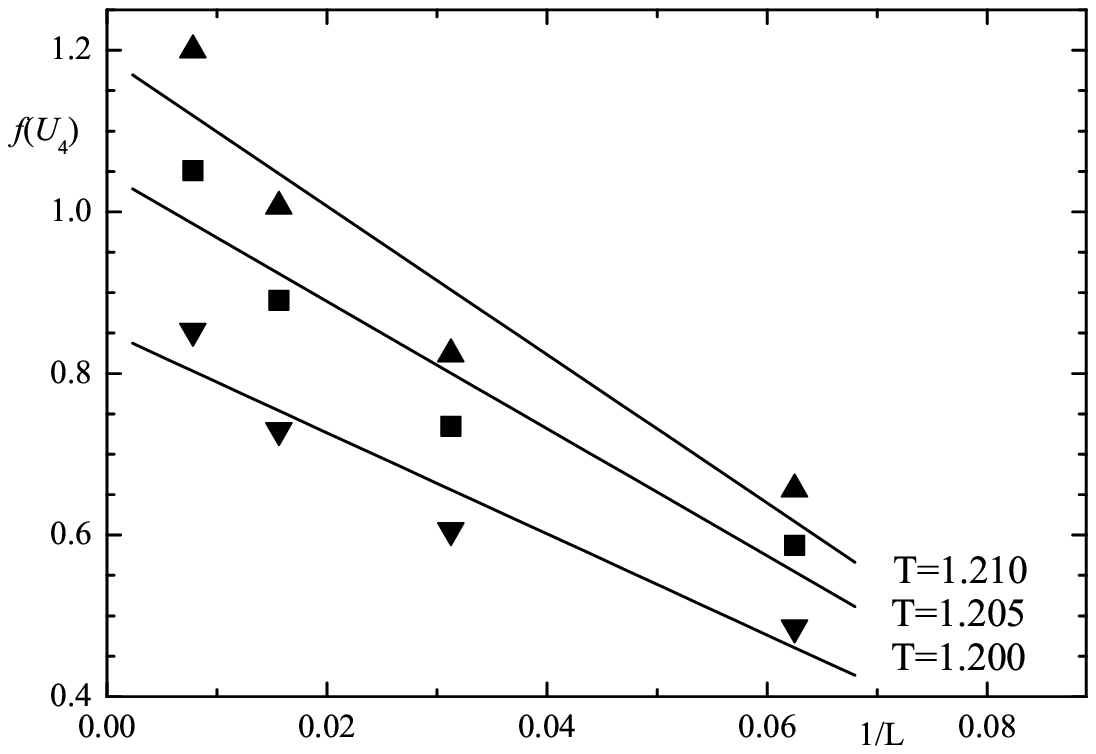}\\[-2mm]
\parbox[t]{0.4\textwidth}{
\caption{\label{fig:U4_scail} The cumulant $U_4$ as a
function of a variable
$(T-T_c)L^{1/\nu}(1+bL^{-\omega})$
\newline ( $\bullet$ is corresponds to $T=1.190$,
\newline \hspace*{1.5mm}$\bigstar$ -- $T=1.195$,
$\blacktriangledown$ -- $T=1.200$, \newline
\hspace*{1.5mm}$\scriptstyle\blacksquare$ --
$T=1.205$, $\blacktriangle$ -- $T=1.210$). }} \quad
\parbox[t]{0.55\textwidth}{
\caption{\label{fig:lnU4} Dependence $f(U_4)=[ \ln U_4/\ln(T-T_c)
]^{-1}$ \newline as a function of $1/L$.}}
\end{figure}
The cumulant $U_4$ as a function of the scaling
variable $(T-T_c)L^{1/\nu}(1+bL^{-\omega})$ with using
correction to scaling procedure for calculated values
$\nu=0.770$ and $\omega=0.786$ is shown in
Fig.~\ref{fig:U4_scail}. Approximation of function
$f(U_4,1/L)=[ \ln U_4/\ln(T-T_c) ]^{-1}$
(Fig.~\ref{fig:lnU4}) gives possibility to estimate
the exponent $\nu=0.758(10)$, which is in a good
agreement with final value $\nu=0.770(74)$
(Table~\ref{tab:all}).

The dynamic and static critical exponents were computed with the use
of the leading corrections to scaling for the 3D Heisenberg model
with linear defects and their values $z=2.257(61)$, $\nu=0.770(74)$,
$\beta=0.393(77)$, and $\omega=0.786(45)$
can be considered as final. In a summary
Table~\ref{tab:all}, we present the values of critical exponents
$z$, $\beta/\nu$, $\nu$, $\beta$, and $\omega$ obtained in this
paper by comprehensive Monte Carlo simulations of the short-time
critical evolution of the diluted 3D Heisenberg model with linear
defects from an ordered initial state with $m_0=1$. For comparison,
we give in Table~\ref{tab:all} the results of renormalization group
(RG) and Monte Carlo (MC) calculations of these exponents for pure
3D Heisenberg model
\cite{GuidaZinnJustin,Prudnikov_JETP2008,FerrenbergLandau} and
diluted 3D Heisenberg model with linear defects
\cite{Prudnikov_PRB2000,Prudnikov_OmSU2010,BlavatskaFerberHol}.

The obtained in this article values of exponents demonstrate very
good agreement in the limits of statistical errors of simulation and
numerical approximations with results of the RG field-theoretic
description from Ref.~\cite{Prudnikov_PRB2000}, calculated with the
use of the Pad\'{e}-Borel (PB) summation technique to $d=3$
expansion series, and particularly with results from
Ref.~\cite{Prudnikov_OmSU2010}, where the Pad\'{e}-Borel-Leroy (PBL)
and the self-similar approximation (SSA) \cite{Yukalov} resummation
methods were also applied to series from
Ref.~\cite{Prudnikov_PRB2000}. The obtained value of the
correction-to-scaling exponent $\omega=0.78(31)$ demonstrates a
sufficiently good agreement with value of $\omega=0.88$, obtained in
Ref.~\cite{BlavatskaFerberHol} by the RG field-theoretical method
with fixed dimension ($d=3$) for Heisenberg model with isotropic
distributed linear defects ($a=2$).

The obtained results confirm the strong influence of LR-correlated
quenched defects on the critical behavior of the systems described
by the many-component order parameter. As a result, a wider class of
disordered systems, not only the three-dimensional diluted Ising
model, can be characterized by a new type of critical behavior
induced by randomly distributed quenched defects and effects of
their spatial correlations.

\section*{Acknowledgements}
The authors would like to thank Prof. V.V. Prudnikov and Dr. A.N. Vakilov
for useful discussion of results.
This work was supported in part by Ministry of Education and Science
of Russia through project No.~02.740.11.0541, by the Russian
Foundation for Basic Research through Grants No.~10-02-00507 and
No.~10-02-00787 and by Grant No.~MK-3815.2010.2 of Russian
Federation President. Our simulations were carried out on the
SKIF-MSU in the Moscow State University and MVS15k in Joint Super
Computer Center of Russian Academy of Sciences.

\end{document}